\documentclass[a4paper,reqno]{amsart}

\usepackage{cite}

\newcommand{\HH}{\mathcal{H}}

\newcommand{\CC}{\mathbb{C}}
\newcommand{\RR}{\mathbb{R}}
\newcommand{\ZZ}{\mathbb{Z}}
\newcommand{\NN}{\mathbb{N}}
\newcommand{\UU}{\mathcal{U}}
\newcommand{\SSS}{\mathbb{S}}

\newcommand{\dom}{\mathop{\mathrm{dom}}}
\newcommand{\diag}{\mathop{\mathrm{diag}}}
\newcommand{\sgn}{\mathop{\mathrm{sign}}}

\newtheorem{prop}{Proposition}
\newtheorem{corol}[prop]{Corollary}

\theoremstyle{definition}

\newtheorem{example}[prop]{Example}

\begin{document}

\title{Reducible boundary conditions in coupled channels}

\author{Konstantin Pankrashkin}

\address{Institut f\"ur Mathematik, Humboldt-Universit\"at zu Berlin,
Rudower Chaussee 25, 12489 Berlin Germany}

\email{const@mathematik.hu-berlin.de}


\begin{abstract}
We study Hamiltonians with point interactions
in spaces of vector-valued functions. Using some information from the theory of quantum graphs we describe a class of the operators
which can be reduced to the direct sum of several one-dimensional problems.
It shown that such reduction is closely connected with the invariance under channel permutations. Examples are provided by some ``model'' interactions, in particular, the so-called $\delta$-, $\delta'$-, and the Kirchhoff couplings.
\end{abstract}


\maketitle

\section{Introduction}

Quantum-mechanical Hamiltonians in coupled channels are a natural generalization
of the Schr\"odinger operators with zero-range interactions~\cite{AGHH}.
They provide a simple example of matrix-valued Fermi pseudopotentials \cite{WUXU}
and can be described using the tools of the extension theory \cite{CNT,CNT2}.
At the same time, such systems can be viewed as special quantum graphs~\cite{ES,gerp,KS,Ku1},
so that one can use the general technique in order to describe all possible interactions in channels,
their spectral characteristics etc. In the present paper, we develop the formalism
of coupled channels using the quantum graph approach.

Such an approach gives a possibility to use the self-adjoint extension theory
and the symplectic technique~\cite{pavlov}
in order to describe all possible boundary conditions. This can be done in many ways
including the transfer matrix formalism, which is widely used in scalar one-dimensional
point interactions~\cite{PS}. Using the representations obtained we show that 
a wide class of the Hamiltonians in question
admits decoupling, i.e. by a certain unitary transformation
one can reduce them to the direct sum of well-studied Schr\"odinger operators with point interactions; we call such Hamiltonians as well as the corresponding boundary conditions \emph{reducible}. Such conditions can be formulated in various terms, including continuity properties of functions from the domain of the Hamiltonian.
An essential feature of the matrix Hamiltonians considered as quantum graphs
is the presence of isometric parts (channels).
In general, we show that the possibility of the reduction to scalar problems is always
connected with certain invariance properties of the boundary conditions with respect to channel permutations. More precisely, it is proved that the matrix Hamiltonian is reducible iff the boundary conditions are invariant under the cyclic coordinate shift in a certain orthonormal basis; similar correspondence was found recently in connection
with the inverse scattering problem on graphs~\cite{BK}.  Although such a decoupling is not a generic property, many ``standard'' boundary conditions appear to be reducible, in particular,  the so-called $\delta$, $\delta_p$, $\delta'$, and $\delta'_s$ couplings as well as Kirchhoff's boundary conditions~\cite{CE} are reducible (in our opinion, this may illustrate the difference between the general quantum graphs and the coupled channels: the model interactions on graphs appear to be trivial in channels, although one can use the same technique for their study). 
The reduction permits us to describe the spectrum
of the simplest matrix Kronig-Penney Hamiltonians (periodically coupled channels)
and to show how their parameters influence various spectral effects like embedded eigenvalues or the number of gaps.

\section{Parameterization of boundary conditions in coupled channels}\label{sec2}

Consider a free particle on a graph, with the Hamiltonian acting
on each edge as $\psi_j\mapsto -\psi''_j$, where $j$ is the edge index.
Assume that the graph has the simplest structure, i.e. consists of $n$ half-lines
$[0,\infty)$ coupled at the origin. The boundary conditions take the form $A\psi(0)=B\psi'(0)$, where
$A$ and $B$ are $n\times n$ matrices satisfying the following two conditions~\cite{KS}:
\begin{subequations}
            \label{BC-KS}
\begin{gather}
AB^*=BA^*, \label{KSa}\\
\text{the block matrix $(AB)$ has maximal rank.}\label{KSb}
\end{gather}
\end{subequations}
The condition~\eqref{KSb} may be rewritten in an equivalent form $\det (AA^*+BB^*)\ne 0$
or $\det (B\pm iA)\ne 0$~\cite{RB}.

Now let us consider the free motion on $n$ lines coupled at some point $q\in\RR$. The Hamiltonian of the problem
is  the operator $H=-d^2/dx^2$ acting in the space $L^2(\RR,\CC^n)$, and the coupling
is described by some boundary conditions at $q$. Formally one can consider the $n$ coupled
lines as $2n$ coupled half-lines, so that all possible boundary conditions take the form
\begin{equation}
          \label{BC-ABff}
A\begin{pmatrix} f(q-)\\
f(q+)
\end{pmatrix}=B\begin{pmatrix} -f'(q-)\\
f'(q+)
\end{pmatrix},
\end{equation}
where $A$ and $B$ are $2n\times 2n$ matrices satisfying \eqref{BC-KS}.
From the other point of view, the nature of coupled channels requests other types
of parameterization~\cite{CNT2}, namely, the transfer matrix formalism, 
\begin{equation}
       \label{BC-trans}
\begin{pmatrix}
f'(q+)\\ f(q+)
\end{pmatrix}
=\begin{pmatrix}
C_{11} & C_{12}\\
C_{21} & C_{22}
\end{pmatrix}
\begin{pmatrix}
f'(q-)\\ f(q-)
\end{pmatrix}.
\end{equation}
Below we will use mainly boundary conditions of the form \eqref{BC-ABff}. Nevertheless,
in many situations it is useful to know the connection between these two types of parameterization. The following proposition generalizes a construction of~\cite{CNT}.
\begin{prop}
The boundary conditions \eqref{BC-trans} define a self-adjoint operator
in $L^2(\RR,\CC^n)$ iff the matrices $C_{jk}$, $j,k=1,2$, obey
\begin{equation}
             \label{BC-CC}
\begin{gathered}
C_{12} C_{11}^*-C_{11}C_{12}^*=0,\quad
C_{21}C_{22}^*-C_{22} C_{21}^*=0,\\
C_{11}C_{22}^*-C_{12}C_{21}^*=E_n.
\end{gathered}
\end{equation}
The conditions \eqref{BC-CC} are equivalent to
\begin{equation}
             \label{BC-CC2}
\begin{gathered}
C_{11}^*C_{21}-C_{21}^*C_{11}=0,\quad C_{12}^*C_{22}-C_{22}^*C_{12}=0,\\
C_{11}^*C_{22}-C_{21}^*C_{12}=E_n
\end{gathered}
\end{equation}
\end{prop}
\begin{proof}
Substituting the equalities
\begin{align*}
\begin{pmatrix}
f'(q+)\\
f(q+)
\end{pmatrix}&=
\begin{pmatrix}
0 & 0\\
0 & E_n
\end{pmatrix}
\begin{pmatrix}
f(q-)\\
f(q+)
\end{pmatrix}+
\begin{pmatrix}
0 & E_n\\
0 & 0
\end{pmatrix}
\begin{pmatrix}
-f'(q-)\\
f'(q+)
\end{pmatrix},\\
\begin{pmatrix}
f'(q-)\\
f(q-)
\end{pmatrix}&=
\begin{pmatrix}
0 & 0\\
E_n & 0
\end{pmatrix}
\begin{pmatrix}
f(q-)\\
f(q+)
\end{pmatrix}-
\begin{pmatrix}
E_n & 0\\
0 & 0
\end{pmatrix}
\begin{pmatrix}
-f'(q-)\\
f'(q+)
\end{pmatrix},
\end{align*}
into \eqref{BC-trans} we obtain 
\begin{equation}
    \label{BC-CCC}
\begin{pmatrix}
C_{12} & 0\\
C_{22} & -E_n
\end{pmatrix}\begin{pmatrix} f(q-)\\f(q+)\end{pmatrix}=
\begin{pmatrix}
C_{11} & E_n\\
C_{21} & 0
\end{pmatrix}\begin{pmatrix} -f'(q-)\\ f'(q+)\end{pmatrix}.
\end{equation}
These boundary conditions define a self-adjoint operator iff the conditions \eqref{BC-KS}
are satisfied. Eq.~\eqref{KSb} holds due to the presence of the blocks with $E_n$, and Eq.~\eqref{KSa} takes the form
\[
\begin{pmatrix}
C_{12} C_{11}^* & C_{12} C_{21}^*\\
C_{22} C_{11}^*-1 & C_{22} C_{21}^*
\end{pmatrix}=
\begin{pmatrix}
C_{11}C_{12}^*& C_{11}C_{22}^*-1\\
C_{21} C_{12}^* & C_{21}C_{22}^*
\end{pmatrix},
\]
which is exactly \eqref{BC-CC}. These conditions means the equality
\[
\begin{pmatrix}
C_{11} & C_{12}\\
C_{21} & C_{22}
\end{pmatrix}\cdot
\begin{pmatrix}
C_{22}^* & -C_{12}^*\\
-C_{21}^* & C_{11}^*
\end{pmatrix}=E_{2n},
\]
which is equivalent to
\[
\begin{pmatrix}
C_{22}^* & -C_{12}^*\\
-C_{21}^* & C_{11}^*
\end{pmatrix}\cdot \begin{pmatrix}
C_{11} & C_{12}\\
C_{21} & C_{22}
\end{pmatrix}
=E_{2n}
\]
and results in~\eqref{BC-CC2}.
\end{proof}
If $n=1$ (i.e. we have just one channel), the conditions \eqref{BC-CC} are well known~\cite{PS}.
The blocks $C_{jk}$ are just complex numbers, and the conditions 
$C_{12}\overline{C_{11}}= C_{11}\overline{C_{12}}$ and
$C_{21}\overline{C_{22}}= C_{22}\overline{C_{21}}$ mean that $\arg C_{11}=\arg C_{12}=:\theta_1$ and
$\arg C_{21}=\arg C_{22}=:\theta_2$. Put $C_{11}=a e^{i\theta_1}$, $C_{12}=b e^{i\theta_1}$,
$C_{21}=c e^{i\theta_2}$, and $C_{22}=d e^{i\theta_2}$, where $a,b,c,d\in\RR$ and $\theta_1,\theta_2\in[0,2\pi)$.
The third condition in \eqref{BC-CC} reads as $(ad -bc)e^{i(\theta_2-\theta_1)}=1$, which means
that $\theta_1=\theta_2=:\theta$, and the boundary conditions take the form
\[
\begin{pmatrix}
f'(q+)\\
f(q+)
\end{pmatrix}=e^{i\theta}
\begin{pmatrix} a & b\\
c & d
\end{pmatrix}
\begin{pmatrix}
f'(q-)\\
f(q-)
\end{pmatrix},
\quad \theta\in [0,2\pi),
\quad a,b,c,d\in\RR,\quad ad-bc=1.
\]

It is reasonable to call boundary conditions admitting the representation \eqref{BC-trans} \emph{connecting}.
Clearly, some boundary conditions are not connecting, for example, the direct sum of the Dirichlet
at $q-$ and $q+$. Below we discuss some less obvious examples.

Let us return to the parameterization~\eqref{BC-ABff}. The use of the values
\begin{equation}
           \label{BC-Gamma}
\Gamma_1f = (f(q-),f(q+)),\quad \Gamma_2f =(-f'(q-),f'(q+))
\end{equation}
has its origin in the theory of self-adjoint extensions of symmetric operators
\cite{pavlov,BG,Ha}. We recall briefly some notions from the theory of abstract boundary values~\cite{gorb}.
Let $S$ be a symmetric operator in a certain Hilbert space with the domain $\dom S$,
$S^*$ be its adjoint with the domain $\dom S^*$.
Let $V$ be some auxiliary Hilbert space,
and $\Gamma_1$, $\Gamma_2$ be linear maps from $\dom S^*$ to $V$ such that
\begin{equation}
            \label{BC-GG}
\langle f ,S^* g\rangle-\langle S^* f,g \rangle=\langle\Gamma_1 f,\Gamma_2 g\rangle-
\langle\Gamma_2 f,\Gamma_1 g\rangle,\quad \text{for any}\quad f,g\in \dom S^*,
\end{equation}
and for any $(v_1,v_2)\in V\times V$ there exists
$f\in\dom S^*$ with $\Gamma_1f=v_1$, $\Gamma_2 f=v_2$.
If $f\in\dom S^*$, the values $\Gamma_1f$ and $\Gamma_2f $
are called \emph{boundary values} of $f$, and the triple $(V,\Gamma_1,\Gamma_2)$
is called a \emph{boundary triple} of the operator $S$. 
Boundary triple exists iff $S$ has equal deficiency indices (i.e. has self-adjoint extensions), and in this case the dimension of $V$ coincides with this deficiency index, see Chapter~3 in~\cite{gorb} for detailed discussion.
If the space $V$ is finite-dimensional (i.e. if the deficiency indices of $S$ are finite), then all self-adjoint extensions of $S$ are
restrictions of $S^*$ on the elements $f\in\dom S^*$ satisfying
abstract boundary conditions $A\Gamma_1 f=B\Gamma_2 f$,
where $A$ and $B$ are matrices satisfying \eqref{BC-KS}.
To obtain a one-to-one parameterization
of the self-adjoint extensions one can normalize $A$ and $B$ by choosing
unitary matrix $U$ with 
\begin{equation}
     \label{BC-iU}
A=1-U,\quad B=i(1+U).
\end{equation}
Unitary $2n\times 2n$ matrices form a $4n^2$-dimensional real manifold, which is exactly the number
of real parameters in the problem.

Let $q\in\RR$ be fixed. Denote by $S$ the operator acting in $L^2(\RR,\CC^n)$ as $-d^2/dx^2$
on functions from $C_0^\infty(\RR\setminus\{q\},\CC^n)$; this operator is symmetric and has deficiency indices $(2n,2n)$. The adjoint operator $S^*$ acts outside $q$ in the same way on functions
from $W^{2,2}(\RR\setminus \{q\},\CC^n)$, so that the usual integration by parts in \eqref{BC-GG}
leads to $V=\CC^{2n}$ and $\Gamma_1$, $\Gamma_2$ in the form \eqref{BC-Gamma}, see \cite{KS}.
The unitary matrix $U$ in \eqref{BC-iU} is particularly useful in approximation problems \cite{CE}, and we will actively use the representation \eqref{BC-iU} for the boundary conditions \eqref{BC-ABff}.
The choice of a boundary triple is not unique: the dimension
of $V$ is invariant, so one can always assume $V=\CC^{2n}$,
and starting with given boundary operators $\Gamma_1$, $\Gamma_2$
one can describe all possible boundary triples by means of suitable linear transformations~\cite{MS}.
From the point of view of spectral problems it may be reasonable to take as a boundary triple for $S$
the set $(V,\Tilde\Gamma_1,\Tilde\Gamma_2)$ with
\begin{equation}
   \label{GG-TT}
\Tilde \Gamma_1 f=\begin{pmatrix}
\Tilde\Gamma_{11}f\\\Tilde\Gamma_{12}f
\end{pmatrix}=\begin{pmatrix}
f'(q-)-f'(q+)\\[\bigskipamount]f(q+)-f(q-)
\end{pmatrix},\quad
\Tilde \Gamma_2 f=\begin{pmatrix}
\Tilde\Gamma_{21}f\\\Tilde\Gamma_{22}f
\end{pmatrix}=\begin{pmatrix}
\dfrac{f(q-)+f(q+)}{2} \\[\bigskipamount] \dfrac{f'(q-)+f'(q+)}{2}
\end{pmatrix},
\end{equation}
so that all possible self-adjoint boundary conditions at $q$ take the form 
\begin{equation}
          \label{BC-LM}
L\Tilde \Gamma_1 f=M \Tilde\Gamma_2 f,
\end{equation}
where $L$, $M$ are matrices satisfying the same conditions as $A$ and $B$ in \eqref{BC-KS},
respectively. Denote the corresponding Hamiltonian by $H^{L,M}$.
If the boundary conditions \eqref{BC-ABff} and \eqref{BC-LM} are equivalent, then
one can choose the corresponding pairs of matrices $(A,B)$  and $(L,M)$ in such a way
that they satisfy
\begin{equation}
         \label{BC-ABLM}
\left\{\begin{aligned}
L&=\frac{1}{2}(AD_1 + B D_2),\\
M&=BD_1-AD_2,\\
D_1&=\begin{pmatrix}
0& -E_n\\
0 & E_n
\end{pmatrix},\\
D_2&=\begin{pmatrix}
E_n& 0\\
E_n&0
\end{pmatrix},
\end{aligned}\right.\quad \text{and}\quad
\left\{\begin{aligned}
A&=LK_2-\frac12MK_1,\\
B&=\frac12 MK_2+LK_1,\\
K_1&=\begin{pmatrix}
E_n & E_n\\
0 &0
\end{pmatrix},\\
K_1&=\begin{pmatrix}
0 & 0\\
-E_n &E_n
\end{pmatrix};
\end{aligned}\right.
\end{equation}
We emphasize that due to the non-uniqueness of the parameterization
this correspondence is not unique.

For the sake of completeness we describe also the resolvents of Hamiltonians in coupled channels, which
is useful in spectral problems.
Let $q_1,\dots,q_m$, $m\in \NN$, be points of $\RR$, $q_1<\dots<q_m$. Consider the operator in $L^2(\RR,\CC^n)$
acting as $-d^2/dx^2$ on functions $f\in W^{2,2}(\RR\setminus\{q_1,\dots,q_m\},\CC^m)$ satisfying
\begin{gather*}
L^{(s)} \Tilde\Gamma^{s}_1 f=M^{(s)}\Tilde\Gamma^{(s)}_2 f,\\
\Tilde \Gamma^{(s)}_1 f=
\begin{pmatrix}
f'(q_s-)-f'(q_s+)\\ f(q_s+)-f(q_s-)
\end{pmatrix},\quad
\Tilde \Gamma^{(s)}_2 f=\frac{1}{2}\,\begin{pmatrix}
f(q_s-)+f(q_s+)\\ f'(q_s-)+f'(q_s+)
\end{pmatrix}, \quad s=1,\dots,m,
\end{gather*}
where for each $s$ the $2n\times 2n$ matrices $L^{(s)}$ and $M^{(s)}$ satisfy
the same conditions as $A$ and $B$ before. Denote by $L$ and $M$ the $2mn\times 2mn$ block matrices
$\diag(L^{(1)},\dots,L^{(m)})$ and $\diag(M^{(1)},\dots,M^{(m)})$, respectively.
The Hamiltonian described will be denoted by $H^{L,M}$. Note that the operator
$H^0\equiv H^{E_n,0}$ is just the free Laplacian. The following proposition
is a variant of the Krein resolvent formula expressed in terms of boundary conditions~\cite{AP}.
\begin{prop}
For $\zeta\in \CC\setminus \RR_+$ denote by $Q(\zeta)$ the $2mn\times 2mn$ matrix consisting
of the $2n\times 2n$ blocks $Q^{(l,s)}(\zeta)$,
\[
Q^{(l,s)}(\zeta)=\dfrac{e^{-\sqrt{-\zeta|q_l-q_s|}}}{2}\,
\begin{pmatrix}
\dfrac{1}{\sqrt{-\zeta}}\,E_n & \sgn (q_l-q_s) E_n\\[\bigskipamount]
-\sgn(q_l-q_s) E_n & -\sqrt{-\zeta}\, E_n
\end{pmatrix},\quad l,s=1,\dots,m; 
\]
here and below we assume that $\sgn 0=0$ and that the square root branch is chosen by the condition $\Im\sqrt{-\zeta}>0$ for $\zeta\in(0,+\infty)$.
If such $\zeta$ is a regular value of $H^{L,M}$, then the matrix
$MQ(\zeta)-L$ is invertible and for any
$f=(f_1,\dots,f_n)\in L^2(\RR,\CC^n)$ the following relation holds:
\begin{multline}
   \label{krein-LM}
(H^{L,M}-\zeta)^{-1}f=(H^0-\zeta)^{-1}f\\{}- 
\sum_{s,l=1}^m\bigg(
\sum_{j,k=1}^n \alpha_{2n(s-1)+j,2n(l-1)+k}(\zeta)\langle g^{(l)}_{\bar\zeta},f_k\rangle g^{(s)}_\zeta e_j\\{}
-\sum_{j,k=1}^n \alpha_{2n(s-1)+j,2n(l-1)+n+k}(\zeta)\langle h^{(l)}_{\bar\zeta},f_k\rangle g^{(s)}_\zeta e_j\\
{}-\sum_{j,k=1}^n \alpha_{2n(s-1)+n+j,2n(l-1)+k}(\zeta)\langle g^{(l)}_{\bar\zeta},f_k\rangle h^{(s)}_\zeta  e_j\\
{} -  \sum_{j,k=1}^n \alpha_{2n(s-1)+n+j,2n(l-1)+n+k}(\zeta)\langle h^{(l)}_{\bar\zeta},f_k\rangle h^{(s)}_\zeta e_j\bigg),
\end{multline}
where the numbers $\alpha_{jk}(\zeta)$, $j,k=1,\dots,2mn$, are the entries of $\big(M Q(\zeta)-L\big)^{-1}M$,
\begin{gather*}
g_\zeta^{(s)}(x)=\frac{1}{2\sqrt{-\zeta}}\,e^{-\sqrt{-\zeta}\,|x-q_s|},\\
h_\zeta^{(s)}(x)=\frac{\sgn (x-q_s)}{2}\,e^{-\sqrt{-\zeta}\,|x-q_s|},\quad s=1,\dots,m,
\end{gather*}
and $e_j$, $j=1,\dots,n$, is the standard basis of $\CC^n$.
\end{prop}

Using \eqref{BC-ABLM} and \eqref{BC-CCC} one can easily express the resolvent
in terms of the parameters $A$ and $B$ in \eqref{BC-ABff} or $C_{jk}$, $j,k=1,2$, in \eqref{BC-trans}.

\section{Decoupling of the single-vertex graph}
To emphasize the specifics of the problems with coupled channels let us return to
the case of $n$ half-lines coupled at the origin. The Hamiltonian of the problem is $-d^2/dx^2$ acting in
$L^2\big((0,\infty),\CC^n\big)\equiv \oplus_{j=1}^n L^2(0,+\infty)$ on functions $f\in W^{2,2}\big((0,\infty),\CC^n\big)$
satisfying the boundary conditions $A f(0)=B f'(0)$
with suitable $A$ and $B$ from \eqref{BC-KS}. We normalize
$A$ and $B$ by choosing them in the form \eqref{BC-iU} with suitable $U\in \UU(n)$;
denote the corresponding Hamiltonian by $H_U$.
Choose $\Theta\in \UU(n)$ such that the matrix
$\Theta^{-1}U\Theta$ is diagonal. Denote by the same letter $\Theta$ the associated
unitary transformation
of $L^2\big((0,\infty),\CC^n\big)$, $(\Theta f)(x)=\Theta f(x)$.
For $x\in (0,\infty)$ there holds
$(\Theta f)''(x)=\Theta f''(x)$.
This means that $\Theta$ reduces the boundary conditions to a direct sum,
\[
\diag(1-e^{i\theta_1},\dots,1-e^{i\theta_n}) g(0)=i\,\diag(1+e^{i\theta_1},\dots,1+e^{i\theta_n}) g'(0),
\qquad g=\Theta f, 
\]
where $e^{i\theta_j}$, $j=1,\dots,n$, are the eigenvalues of $U$. In other words,
the operator $H_U$ appears to be unitarily equivalent to the direct sum $\oplus_{j=1}^n H_j$, where
each $H_j$ is a self-adjoint operator in $L^2(0,\infty)$ acting as $-d^2/dx^2$ on functions
$g_j\in W^{2,2}(0,\infty)$ satisfying the boundary conditions
\[
(1-e^{i\theta_j}) g_j(0)=i(1+e^{i\theta_j}) g'_j(0),\quad j=1,\dots,n.
\]
Therefore, the spectral properties of $H_U$ are
determined by the eigenvalues of $U$ only, i.e.
by $n$ parameters from $\SSS^1$; moreover, two such $n$-tuples differing
only by the order of terms are equivalent. This means, in particular,
that the inverse scattering problem on the single vertex graph
has multiple solution; note that matrix $U$
can be still uniquely recovered from the scattering data \cite{KS}.

Although the above schema gives a complete result, its applicability is rather restricted
if the graph contains more than one vertex. One can find some generalizations for ``star-shaped'' graphs,
i.e. if instead of the Hilbert space $\oplus L^2(0,\infty)$ one deals with the space $\oplus L^2(G)$, where
$G$ is some graph; this models $n$ identical graphs $G$ coupled at a certain point.
But even in this case the transformation $\Theta$ mentioned above is non-local
and leads in general to non-local boundary conditions at other vertices of the partial graph $G$.
To obtain a reasonable gain from such a procedure one should consider only diagonalizing transformations
preserving the structure of $G$. We illustrate such a possibility by problems with coupled channels.

\section{Decoupling of channels}

Denote $\HH=\oplus_{j=1}^n L^2(\RR)\equiv L^2(\RR,\CC^n)$. Let $Q$ be a uniformly discrete subset of $\RR$, i.e.
\[\inf_{\substack{p,q\in Q,\\p\ne q}}|p-q|=d>0.\]
On the domain $\dom S=C_0^\infty(\RR\setminus Q,\CC^n)$ consider the operator
$S=-d^2/dx^2$; the adjoint operator $S^*$ acts in the same way on the domain
$\dom S^*=W^{2,2}(\RR\setminus Q,\CC^n)$.
To obtain self-adjoint operators one should introduce boundary conditions
at all points of $Q$ as described in Section~\ref{sec2}; the uniform discreteness of $Q$
guarantees that the operator obtained is self-adjoint \cite{Ku1}.
Such an operator can be interpreted as the Hamiltonian of a free
particle in $n$ channels coupled at the points of $Q$.
We say that such a Hamiltonian $H$ is \emph{reducible} iff
there exists $\Theta\in \UU(n)$ such that the unitary transformation
$\HH\ni f\mapsto \Theta f\in \HH$ reduces $H$ to a direct sum of $n$ one-dimensional
point interaction Hamiltonians. 

All possible boundary conditions at $q\in Q$ have the form
\begin{equation}
        \label{BC-U}
\begin{gathered}
\big(1-U(q)\big)\Gamma_1 f =i\big(1+U(q)\big)\Gamma_2 f \Longleftrightarrow (\Gamma_1-i\Gamma_2)f=U(q)(\Gamma_1+i\Gamma_2)f,\\
\Gamma_1 f =\big(f(q-),f(q+)\big),\quad
\Gamma_2 f =\big(-f'(q-),f(q+)\big), \quad U(q)\in \UU(2n),
\end{gathered}
\end{equation}
Denote the Hamiltonian corresponding to these boundary conditions by $H_{Q,U}$.
Clearly, in the case of finite $Q$ all possible Hamiltonians with point interactions
are parameterized by $4n^2|Q|$ real parameters.

Representing $U(q)$ in the block form,
\[
U(q)=\begin{pmatrix}
U_{11}(q) & U_{12}(q)\\
U_{21}(q) & U_{22}(q)
\end{pmatrix},
\]
we conclude that $H_{Q,U}$ is reducible if and only if the $n\times n$ blocks $U_{jk}(q)$, $j,k=1,2$, $q\in Q$, can be diagonalized simultaneously in some orthogonal basis, i.e. if there exists $\Theta\in \UU(n)$ with
\begin{equation}
      \label{Th-diag}
\begin{gathered}
\begin{pmatrix}
\Theta^{-1}& 0\\
0 & \Theta^{-1}
\end{pmatrix}
\begin{pmatrix}
U_{11}(q) & U_{12}(q)\\
U_{21}(q) & U_{22}(q)
\end{pmatrix}
\begin{pmatrix}
\Theta& 0\\
0 & \Theta
\end{pmatrix}=
\begin{pmatrix}
\Lambda_{11}(q) & \Lambda_{12}(q)\\
\Lambda_{21}(q) & \Lambda_{22}(q)
\end{pmatrix},\\
\Lambda_{jk}(q)=\diag\big(\lambda_{jk}(q,s)\big),\\
\lambda_{jk}(q,s) \text{ are the eigenvalues of }U_{jk}(q),\quad
s=1,\dots,n,\quad j,k=1,2,\quad q\in Q.
\end{gathered}
\end{equation}
The unitary transformation $\HH\ni f\mapsto\Theta f \in \HH$
reduces $H_{Q,U}$ to the direct sum of one-dimensional Hamiltonians, as
due to \eqref{Th-diag} and to the equalities $(\Theta f)(q\pm)=\Theta \big(f(q\pm)\big)$ and $(\Theta f)'(q\pm)=\Theta \big(f'(q\pm)\big)$ the boundary conditions \eqref{BC-U} for $g=\Theta f$ take the form
\begin{equation}
\begin{pmatrix}
g_s(q-)+ig'_s(q-)\\
g_s(q+)-ig'_s(q+)
\end{pmatrix}=\begin{pmatrix}
\lambda_{11}(q,s) &\lambda_{12}(q,s)\\
\lambda_{21}(q,s) &\lambda_{22}(q,s) 
\end{pmatrix}\begin{pmatrix}
g_s(q-)-ig'_s(q-)\\
g_s(q+)+ig'_s(q+)
\end{pmatrix},
\quad s=1,\dots,n.
\end{equation}  
For the generic interaction, the condition for the reducibility of the boundary conditions can be formulated as follows:

\begin{prop} \label{prop-ujk}
The boundary conditions \eqref{BC-U} are reducible if and only if all
the blocks $U_{jk}(q)$, $j,k=1,2$, $q\in Q$, are normal and commute with each other.
\end{prop}
This means that the reducible boundary conditions are parameterized, roughly speaking,
by $n|Q|$ unitary $2\times 2$ matrices  $\big(\lambda_{jk}(q,s)\big)_{j,k=1,2}$, $q\in Q$, $s=1,\dots,n$ (up to permutations),
and a unitary $n\times n$ matrix $\Theta$ which diagonalizes the boundary conditions.
This means that the most general reducible boundary conditions involve
$n|Q|\dim_\RR \UU(2)+\dim_\RR\UU(n)=4n|Q|+n^2$ real parameters.

An analogue of Proposition~\ref{prop-ujk}
can be given in terms of the transfer matrix~\eqref{BC-trans}.
\begin{prop}
The Hamiltonian $H_{Q,U}$ given by the boundary conditions~\eqref{BC-trans} 
is reducible iff the blocks $C_{jk}$, $j,k=1,2$, are normal and commute
for all $q\in Q$. In particular, if $Q$ consists of a single point $q$ and
$C_{jk}$ are self-adjoint, then $H_{Q,U}$ is reducible. 
\end{prop}
\begin{proof}
The first part is obvious. Assuming that that the blocks $C_{jk}$, $j,k=1,2$, are self-adjoint (like it was done in~\cite{CNT2}),
one concludes from \eqref{BC-CC} and \eqref{BC-CC2}
that they all commute with each other and, therefore, can be diagonalized simultaneously.
But this means that the corresponding Hamiltonian is reducible.
\end{proof}

It is useful also to have ``quantitative'' reducibility criteria in terms of boundary conditions.
The corresponding matrix $U$ may be difficult to find, but the reducibility can be found by other means.
To illustrate this, we consider the Hamiltonian $\Tilde H$
given by its quadratic form $\Tilde Q (f,f)=\langle f',f'\rangle+\langle f(q),Af(q)\rangle$, $\dom \Tilde Q=W^{1,2}(\RR,\CC^n)$, where $A$ is a $n\times n$ self-adjoint matrix. This Hamiltonian may be viewed
as the so-called matrix $\delta$-potential and corresponds to the boundary conditions
$f(q-)=f(q+)=:f(q)$, $f'(q+)-f'(q-)=A f(q)$, cf.~\cite{Ku1}. Clearly, an orthogonal transformation which diagonalizes $A$ will reduce the boundary conditions to a direct sum; the Hamiltonian $\Tilde H$ is unitarily equivalent
to the operator $-d^2/dx^2$ with the boundary conditions
$g_j(q-)=g_j(q+)=:g_j(q)$, $g'_j(q+)-g'_j(q-)=\alpha_j g_j(q)$, $j=1,\dots,n$,
respectively, where $\alpha_j$ are the eigenvalues of $A$. In other words, $\Tilde H$ is
isomorph to the direct sum of the usual one-dimensional $\delta$-perturbations.
Let us try to generalize this example.
\begin{prop}\label{prop3}
Let $Q$ consist of a single point $q$.
If there exist $\alpha,\beta\in\RR$, $|\alpha|+|\beta|>0$, and $c,c'\in\{-1,1\}$ such that
\begin{equation}
          \label{BC-fab}
\alpha\big(f'(q+)+c' f'(q-)\big)=\beta\big(f(q+)+cf(q-)\big)
\end{equation} for all $f\in \dom H_U$,
then $H_U\equiv H_{Q,U}$ is reducible.
\end{prop}

\begin{proof}
Consider first the case $c=c'=-1$. 
Assume first $\alpha=0$. Put $D=\{f\in\dom S^*:\, f(q-)=f(q+)=:f(q)\}$.
Clearly, $\dom H_U\subset D$, and for arbitrary $f,g\in D$ there holds
\begin{multline*}
\langle \Gamma_1 f,\Gamma_2 g\rangle-\langle \Gamma_2 f,\Gamma_1 g\rangle
{}\equiv-\langle f(q-),g'(q-)\rangle+\langle f(q+),g'(q+)\\{}+\langle f'(q-),g(q-)\rangle-\langle f'(q+),g(q+)\rangle
=\langle\Gamma'_1 f, \Gamma'_2 g\rangle-\langle \Gamma'_2f,\Gamma'_1 g\rangle,
\end{multline*}
where $\Gamma'_1 f=f(q)$, $\Gamma'_2 f=f'(q+)-f'(q-)$. Denote by $S_0$ the restriction
of $S^*$ to the set $\dom S_0=\{f\in D:\,\Gamma'_1f=\Gamma'_2 f=0\}\equiv \{f\in\dom S^*:\,f(q-)=f(q+)=0,\,
f'(q+)=f'(q-)\}$. Clearly, this is a symmetric operator, and the set $D$ is the domain
of its adjoint $S_0^*$. Therefore, $(\CC^n,\Gamma'_1$, $\Gamma'_2)$ is
 a boundary triple for this new operator
$S_0$. As $H_U$ is a self-adjoint extension of $S_0$, there exists
$V\in \UU(n)$ so that $H_U$ is determined by the boundary conditions $(\Gamma'_1-i\Gamma'_2)f=V(\Gamma'_1+i\Gamma'_2)f$,
$f\in\dom S_0^*$. Let $\Theta$ be a unitary transformation which diagonalizes $V$.
Clearly, $\Theta$ induces a unitary transformation of $\HH$, and the components
of the function $g=\Theta f$, $f\in\dom H_U$, satisfy
\begin{gather}
g_j(q-)=g_j(q+)=:g(q),\quad (1-e^{i\theta_j})g_j(q)=i(1+e^{i\theta_j})\cdot \big(g'_j(q+)-g'_j(q-)\big),\\
e^{i\theta_j} \text{ are eigenvalues of } V, \quad j=1,\dots,n.
\end{gather}
Therefore, $H_U$ is reducible.

Consider now the case $\alpha\ne 0$. Put $\gamma=\beta/\alpha$.
We use the boundary triple \eqref{GG-TT}. Denote by $D$ the set $\{f\in\dom S^*:\,\Tilde\Gamma_{11}f-\gamma\Tilde \Gamma_{12}f=0\}$. The condition \eqref{BC-fab} means the inclusion $\dom H_U\subset D$.
Let $f,g\in D$, then
$\langle \Tilde\Gamma_1 f,\Tilde\Gamma_2 g\rangle-\langle\Tilde\Gamma_2 f,\Tilde\Gamma_1 g\rangle=\langle\Gamma'_1 f,\Gamma'_2g\rangle-\langle \Gamma'_2 f,\Gamma'_1 g\rangle$ with
$\Gamma'_1 f=\Tilde \Gamma_{12} f$, $\Gamma'_2 f=\gamma\Tilde\Gamma_{21}f+\Tilde\Gamma_{22}f$.
Denote by $S_0$ the restriction of $S^*$ to the set $\dom S_0=\{f\in D:\, \Gamma'_1 f=\Gamma'_2 f=0\}$;
this is a symmetric operator, $D=\dom S_0^*$, and $(\CC^n,\Gamma'_1,\Gamma'_2)$ is a boundary triple for $S_0$.
As $H_U$ is a self-adjoint extension of $S_0$, there exists $V\in \UU(n)$ such that
$H_U$ is determined by the boundary conditions
 $(\Gamma'_1-i\Gamma'_2)f=V(\Gamma'_1+i\Gamma'_2)f$, $f\in D$.
Let $\Theta\in \UU(n)$ such that $\Theta^{-1}V\Theta$ is diagonal. Noting that
$\Theta$ commutes with all the operators $\Tilde \Gamma_{jk}$, $\Gamma'_j$, $j,k=1,2$,
we reduce the boundary conditions to a direct sum for $g=\Theta f$.

Now let $c'=-1$, $c=1$. Denote by $D$ the set of functions $f\in\dom S^*$ satisfying \eqref{BC-fab}
and use again the boundary triple  \eqref{GG-TT}, then for any $f,g\in D$ there holds
$\langle\Tilde\Gamma_1f,\Tilde \Gamma_2 g\rangle-\langle\Tilde\Gamma_2f,\Tilde \Gamma_1 g\rangle=
\langle \Gamma'_1 f,\Gamma'_2g\rangle-\langle\Gamma'_2 f,\Gamma'_1g\rangle$ with
$\Gamma'_1 f=\Tilde\Gamma_{12}f$, $\Gamma'_2 f=\Tilde\Gamma_{22}f$. Denote by $S_0$ the symmetric operator
which is the restriction of $S^*$ to the domain $\dom S_0=\{f\in D:\,\Gamma'_1f=\Gamma'_2f=0\}$,
then $D=\dom S_0^*$. Taking into account the fact that $H_U$ is a self-adjoint
extension of $S_0$ we proceed with the proof as in the previous case.
The rest combinations of $c$ and $c'$ can be considered in the same way.
\end{proof}

To formulate an important corollary we recall that a function
$f:\RR\to\RR$ is called anticontinuous at $q$ if there exist
the limits $f(q\pm)$ and $f(q+)+f(q-)=0$.
\begin{corol}
Let the set $Q$ consist of a single point $q$. If one of the following conditions is satisfied:
\begin{itemize}
\item all functions from $\dom H_{Q,U}$ are continuous,
\item all functions from $\dom H_{Q,U}$ are anticontinuous,
\item derivatives of all functions from $\dom H_{Q,U}$ are continuous,
\item derivatives of all functions from $\dom H_{Q,U}$ are anticontinuous,
\end{itemize}
then $H_{Q,U}$ is reducible.
\end{corol}
We emphasize again that the last proposition and the corollary apply to channels coupled at one point only. Of course, this works also for channels which are
identically coupled at several points.

\section{Permutation-invariant boundary conditions}\label{sec-ddp}

Let us return to the general Hamiltonian $H_{Q,U}$ with boundary conditions at point
of a discrete set $Q$ (see the beginning of the previous section). The aim of this section is to discuss a correspondence between the reducibility and invariance under channel permutations.

The most general version of this correspondence can be formulated as follows:
\begin{prop}\label{pinv}
A matrix-valued point interaction Hamiltonian $H_{Q,U}$ with a point interaction supported
by a uniformly discrete set $Q$ is reducible if and only if there exists
a unitary $n\times n$ matrix $\Theta$ with non-degenerate eigenvalues
such that the boundary conditions
at all points of $Q$ are invariant under the transformation
$f\mapsto\Theta f$.
\end{prop}

\begin{proof}
At each point $q\in Q$ there exists $U=U(q)\in \UU(2n)$ such that all the functions
from the domain of $H_{Q,U}$ are characterized by the condition~\eqref{BC-U}. 

If the Hamiltonian is reducible, all the blocks $U_{jk}(q)$, $j,k=1,2$, $q\in Q$, 
are diagonal in some orthogonal basis and. therefore, commute with any
matrix which is diagonal in this basis. Taking an arbitrary diagonal
unitary matrix with non-degenerate eigenvalues we show that the condition
formulated in the proposition is necessary. Let us show that this condition is also sufficient.

The invariance of boundary conditions under $\Theta$ means 
that all the blocks $U_{jk}(q)$, $j,k=1,\dots,n$,
commute with $\Theta$. This means that the invariant subspaces of $\Theta$ are such for \emph{all} blocks
at \emph{all} points of $Q$. As these subspaces are one-dimensional and orthogonal to each other, all the blocks are diagonal in the eigenbasis of $\Theta$.
\end{proof}

This proposition shows that the reducibility is an effect which is closely connected
with non-uniqueness in the inverse scattering or spectral problems on graphs~\cite{BK}:
If the Hamiltonian is invariant under a unitary transformation
with certain properties, then there exists another graph (in our case, the union of real lines with marked points) having the same spectrum.

An important example is provided by Hamiltonians which are invariant under certain channel permutations.

\begin{corol}
Let $\sigma$ be a permutation of order $n$ (i.e. $\sigma^n=\text{id}$ and $\sigma^k\ne\text{id}$ for all $k\in\{1,\dots,n-1\}$)
and $H_{Q,U}$ be invariant under the transformation $f_j\mapsto f_{\sigma(j)}$, $j=1,\dots,n$, then $H_{Q,U}$ is reducible.
\end{corol}
\begin{proof}
Indeed, the minimal polynom of the transformation is $\lambda^n-1$, which means that
all eigenvalues are simple.
\end{proof}
Actually, this situation is in some sense generic, as the following proposition shows.
\begin{prop}
The Hamiltonian $H_{Q,U}$ is reducible iff there exists an orthonormal basis $(h_1,\dots,h_n)$ in $\CC^n$, so that all boundary conditions are invariant under the transformation $h_{j}\mapsto h_{(j-1)\,\mathrm{mod}\,n}$.
\end{prop}
\begin{proof}
The minimal polynom of the transformation described is again $\lambda^n-1$, which means that all eigenvalues are simple. This shows that the existence of such transformation
is sufficient for the boundary conditions to be reducible. Let us show that this condition is also necessary.

Let $H_{Q,U}$ be reducible, then there exists an orthonormal basis $G=(g_j, j=1,\dots,n)$ in $\CC^n$ in which all blocks $U_{jk}(q)$ are diagonal.
In this basis,
define a linear transformation $\Xi$ by its matrix $\diag(\lambda_1,\dots,\lambda_n)$,
$\lambda_j=\exp(2\pi i j/n)$, $j=1,\dots,n$. Clearly, the blocks
$U_{jk}(q)$ commute with $\Xi$ (as all these matrices are diagonal).
From the other side, in the basis
\[
h_j=\dfrac{1}{\sqrt{n}}\,\sum_{k=1}^n \bar\lambda^k_j g_k,\quad j=1,\dots,n,
\]
the transformation $\Xi$ has the matrix
\[
\begin{pmatrix}
0 & 1 & 0& \dots &0\\
0 & 0 & 1& \dots &0\\
\dots &\dots&\dots &\dots &\dots\\
0 &0 & 0 & \dots &1\\
1 &0 & 0 & \dots &0
\end{pmatrix},
\]
which is exactly the matrix of the transformation
\[
\sum_{j=1}^n\langle h_j,f\rangle h_j\mapsto
\sum_{j=2}^n\langle h_j,f\rangle h_{j-1}+\langle h_1,f\rangle h_n.
\]
Therefore, the boundary conditions at all points are invariant under the cyclic shift of coordinates with respect to the basis $(h_1,\dots,h_n)$.
\end{proof}

Of certain interest are Hamiltonians (and the corresponding boundary conditions)
which are invariant under \emph{all} channel permutations.
Clearly, this means that the blocks $U_{jk}(q)$ are of the form
$U_{jk}(q)=a_{jk}(q)E_n+b_{jk}(q)J_n$, where $J_n$ is the $n\times n$ matrix whose all entries are equal to $1$, and the complex numbers $a_{jk}(q)$ and $b_{jk}(q)$ obey the condition
$\big(a_{jk}(q)\big)_{j,k=1,2},\big(a_{jk}(q)+n b_{jk}(q)\big)_{j,k=1,2}\in \UU(2)$.
(Clearly, the spectrum of $J_n$ consists of a simple eigenvalue $n$ and a $(n-1)$-fold degenerate eigenvalue $0$.) This class includes the frequently used
$\delta$, $\delta'_s$, $\delta_p$, and $\delta'$ couplings, which we consider in
greater detail (some different notation is used, see~\cite{WUXU}).
For more detailed discussion of the origin of these coupling types we refer
to the works \cite{CE,Ex96} and references therein.
The corresponding  boundary conditions for a function
$f\in W^{2,2}(\RR\setminus\{q\},\CC^n)$ are as follows:
\begin{subequations}
       \label{BC-D}
\begin{align}
        \label{BC-delta}
\delta(q,\alpha):\quad& \left\{\begin{aligned}f_j(q-)=f_k(q+)&=:f(q),\quad j,k=1,\dots,n,\\
\sum_{j=1}^n \big(f'_j(q+)-f'_j(q-)\big)&=\alpha f(q),
\end{aligned}\right.\\
        \label{BC-deltas}
\delta'_s(q,\beta):\quad & \left\{\begin{aligned}-f_j'(q-)=f'_k(q+)&=:f'(q),\quad j,k=1,\dots,n,\\
\sum_{j=1}^n \big(f_j(q+)+f_j(q-)\big)&=\beta f'(q),
\end{aligned}\right.\\
        \label{BC-deltap}
\delta_p(q,\alpha):\quad& \left\{\begin{aligned}
\pm f'_j(q\pm)\mp f'_k(q\pm)&=\frac{\alpha}{2n}\big(f_j(q\pm)-f_k(q\pm)\big),\quad
j,k=1,\dots,n,\\
\sum_{j=1}^n \big(f_j(q-)+f_j(q+)\big)&=0,
\end{aligned}\right.\\
        \label{BC-deltaprim}
\delta'(q,\beta):\quad& \left\{\begin{aligned}
f_j(q\pm)-f_k(q\pm)&=\frac{\beta}{2n}\big(\pm f'_j(q\pm)\mp f'_k(q\pm)\big),\quad
j,k=1,\dots,n,\\
\sum_{j=1}^n \big(f'_j(q+)-f'_j(q-)\big)&=0,
\end{aligned}\right.
\end{align}
\end{subequations}
where $\alpha$ and $\beta$ are real parameters.
The $\delta(q,0)$-coupling corresponds to the so-called \emph{Kirchhoff boundary conditions} at $q$; they appear, for example, if one considers the coupled channels as a limit of shrinking manifolds~\cite{EP}.
For the sake of brevity we denote the introducing of boundary conditions as 
a formal sum, for example, under the operator 
\begin{equation}
          \label{H-sum}
H=-\frac{d^2}{dx^2}+\delta'_s(q_1,\beta)+\delta(q_2,\alpha)
\end{equation}
we mean the operator which acts as $f\mapsto -f''$ on functions $f\in W^{2,2}(\RR\setminus\{q_1,q_2\},\CC^n)$
satisfying the boundary condition \eqref{BC-deltas} for $q=q_1$ and \eqref{BC-delta} for $q=q_2$. In one-dimensional case we
use a more traditional
way of writing, for example,
\begin{equation}
         \label{H-sum2}
H=-\frac{d^2}{dx^2}+\beta\delta'_s(x-q_1)+\alpha\delta(\alpha-q_2)
\end{equation}
will denote the same operator as in \eqref{H-sum}
\emph{assuming that $n=1$}. In fact, one can consider the expression
\eqref{H-sum2} as a self-adjoint operator if one uses the theory of distributions with discontinuous test functions~\cite{AK}, see
also~\cite{WUXU}.

\begin{prop}\label{prop2}
Let $Q$, $Q'_s$, $Q_p$, $Q'$ be non-intersecting discrete subsets of $\RR$,
and their union $P:=Q\cup Q'_s\cup Q_p\cup Q'$ be uniformly discrete.
Denote by $H$ the self-adjoint operator in $L^2(\RR,\CC^n)$, $n>1$, of the form
\[
-\frac{d^2}{dx^2}+\sum_{q\in Q}\delta(q,\alpha_q)+\sum_{q\in Q'_s}\delta'_s(q,\beta_q)+\sum_{q\in Q_p}\delta_p(q,\alpha_q)
+\sum_{q\in Q'}\delta'(q,\beta_q),
\]
where $\alpha_q$, $\beta_q$ are real parameters.
Then $H$ is unitarily equivalent to the direct sum $\oplus_{k=1}^n H_k$, where
$H_k$ are self-adjoint operators in $L^2(\RR)$, namely,
\begin{equation}
      \label{BC-h1}
H_1=-\frac{d^2}{dx^2}+\sum_{q\in Q}\frac{\alpha_q}{n}\,\delta(x-q)+\sum_{q\in Q'_s}\frac{\beta_q}{n}\,\delta'_s(x-q)+\sum_{q\in Q_p}\frac{\alpha_q}{n}\,\delta_p(x-q)
+\sum_{q\in Q'}\frac{\beta_q}{n} \delta'(x-q),
\end{equation}
i.e. the operator $-d^2/dx^2$ acting on functions $f\in W^{2,2}(\RR\setminus P)$ satisfying
\begin{align*}
f(q-)=f(q+)&=:f(q),&\quad f'(q+)-f'(q-)&=\frac{\alpha_q}{n}\,f(q), & q\in Q, \\
f'(q-)+f'(q+)&=0, & \quad f(q-)+f(q+)&=\frac{\beta_q}{n}\,f(q+), & q\in Q'_s,\\
f(q-)+f(q+)&=0, & \quad f'(q+)-f'(q-)&=\frac{\alpha_q}{n}\,f'(q+), & q\in Q_p,\\
f'(q-)=f'(q+)&=:f'(q), & \quad f(q+)-f(q-)&=\frac{\beta_q}{n}\,f'(q), & q\in Q'.
\end{align*}
and the operators $H_2,\dots, H_n$ are equal to each other and act as $g(x)\mapsto -g''(x)$, $x\notin P$,
on functions $g\in W^{2,2}(\RR\setminus P)$ satisfying the following boundary conditions:
\begin{equation}
       \label{BC-sep}
\begin{gathered}
g(q-)=g(q+)=0, \quad q\in Q,\qquad
g'(q-)=g'(q+)=0, \quad q\in Q'_s,\\
\alpha_q g(q-)+2n g'(q-)=\alpha_q g(q+)-2n g'(q+)=0, \quad q\in Q_p,\\
2n g(q-)+\beta_q g'(q-)=2n g(q+)-\beta_q g'(q+)=0,\quad q\in Q'.
\end{gathered}
\end{equation}
\end{prop}
 
\begin{proof} We recall that the uniform discreteness of $P$ guarantees
the self-adjointness of $H$ \cite{Ku1}.

As it was shown in \cite{CE}, the boundary conditions \eqref{BC-D}
can be written as \eqref{BC-U} with $U(q)=a_q E_{2n}+b_q J_{2n}$, where
\begin{equation}
a_q=\begin{cases}
-1& \text{for }\delta(q,\alpha_q),\\[\medskipamount]
1&\text{for }\delta'_s(q,\beta_q),\\[\medskipamount]
\dfrac{2n-i\alpha_q}{2n+i\alpha_q}&\text{for }\delta_p(q,\alpha_q),\\[\medskipamount]
-\dfrac{2n+i\beta_q}{2n-i\beta_q}&\text{for }\delta'(q,\beta_q),
\end{cases}
\quad
b_q=\begin{cases}
\dfrac{2}{2n+i\alpha_q}  & \text{for }\delta(q,\alpha_q),\\[\medskipamount]
-\dfrac{2}{2n-i\beta_q}  &\text{for }\delta'_s(q,\beta_q),\\[\medskipamount]
-\dfrac{2}{2n+i\alpha_q} &\text{for }\delta_p(q,\alpha_q),\\[\medskipamount]
\dfrac{2}{2n-i\beta_q}\,J_{2n}&\text{for }\delta'(q,\beta_q).
\end{cases}
\end{equation}
The $n\times n$ blocks of $U$ are of a rather simple form, namely,
$U_{11}(q)=U_{22}(q)=a_q E_n+b_q J_n$, $U_{12}(q)=U_{21}(q)=b_q J_n$.
Let $\Xi$ be a linear transformation which diagonalizes $J_n$, then
at each point $q\in P$ the components of the functions $g:=\Xi f$, $f\in\dom H$, satisfy
\begin{equation}
           \label{BC-gU}
\begin{pmatrix}
g_k(q-)+ig'_k(q-)\\
g_k(q+)-ig'_k(q+)
\end{pmatrix}= V_k(q)
\begin{pmatrix}
g_k(q-)-ig'_k(q-)\\
g_k(g+)+ig'_k(q+)
\end{pmatrix},\quad k=1,\dots,n
\end{equation}
with
\[
V_1(q)=\begin{pmatrix}
a_q+nb_q & n b_q\\
nb_q & a_q+n b_q
\end{pmatrix},\qquad
V_k=\begin{pmatrix}
a_q & 0\\
0 & a_q
\end{pmatrix},
\quad k=2,\dots,n,
\]
which is exactly~\eqref{BC-h1} and~\eqref{BC-sep}.
\end{proof}
All the boundary conditions~\eqref{BC-sep} are obviously non-connecting; this means that
none of the couplings~\eqref{BC-D} admits the representation~\eqref{BC-trans}.
This is connected with the fact that these couplings are actually invariant
also under half-channel permutation.

\section{Periodically coupled channels}

Let us illustrate the separability effects by periodic problems with point interactions.
Periodically coupled channels provide simple examples
of periodic quantum graphs, so that the general powerful technique for their analysis
is available~\cite{Ku2,OPS}. The previous discussion gives a possibility
to describe the spectrum of some periodic Hamiltonians by other means:
one can easily reduce the spectral problem for periodically coupled channels
to the spectral problem for periodic scalar Hamiltonians with point interactions, i.e.
to the well-studied generalized Kronig-Penney models~\cite{RH}.
We restrict ourselves by considering some examples.

\begin{example}[Permutation-invariant delta-potential] In $L^2(\RR,\CC^2)$ consider the periodic delta-potential invariant under channel permutation; this corresponds
to the boundary conditions
\[
f(q-)=f(q+)=:f(q),\quad
f'(q+)-f'(q-)=(\alpha E_2+\beta J_2) f(q),\quad
\alpha,\beta\in\RR,\quad q\in \pi\ZZ.
\]
Elementary considerations show that this Hamiltonian $H$ is unitarily equivalent to the direct
sum $H_1\oplus H_2$,
\[
H_1=-\frac{d^2}{dx^2}+\alpha\sum_{n\in\ZZ}\delta(x-\pi n),\quad
H_2=-\frac{d^2}{dx^2}+(\alpha+2\beta)\sum_{n\in\ZZ}\delta(x-\pi n),
\]
so that the spectrum of $H$ is the union of the spectra
of $H_1$ and $H_2$.
If both $\alpha$ and $\alpha+2\beta$ have the same sign, then the spectrum
of $H$ has an infinite number of gaps. For example, for $\alpha,\alpha+2\beta>0$
the spectra of $H_1$ and $H_2$ consist of the bands
$(a_m,m^2)$ and $(b_m,m^2)$, $m=1,2,\dots$, respectively, where
$a_m,b_m>(m-1)^2$, see Theorem~III.2.2.3 in~\cite{AGHH}. The spectrum
of $H$ consists then of the bands $(\min(a_m,b_m),m^2)$, $m=1,2,\dots$.

Let us show that $H$ has only a finite number of gaps
if $\alpha(\alpha+2\beta)<0$.
To be definite, assume that $\alpha>0$ and $\alpha+2\beta<0$ (the second case can be considered in the same way).
The spectrum of $H_1$ consists of the bands
$(a_m,m^2)$, $m=1,2,\dots$, where $a_m=m^2-2m-2\alpha/\pi-1+O(1/m)$, $m\to\infty$,
and the spectrum of $H_2$ consists of the bands $(A_m,B_m)$, $m=1,2,\dots$,
where $A_1<B_1<0$, $B_m>A_m=(m-1)^2$, $m=2,3,\dots$, $B_m=(m-1)^2+2m+2(\alpha+2\beta)-1+O(1/m)$, $m\to\infty$, see Theorem~III.2.2.3 in~\cite{AGHH}. Obviously, for large $m$ there holds $B_m>a_m$, which means
that the gaps are overlapped by the large bands. 
\end{example}

\begin{example}[Periodic $\delta$-coupling] \label{pdc}
For any real $\alpha$ the operator
\[
H=-\frac{d^2}{dx^2}+\sum_{l\in\ZZ}\delta(\pi l,\alpha)
\]
acting in $L^2(\RR,\CC^n)$ is unitarily equivalent to the direct sum
$H_\alpha\oplus(\oplus_{j=1}^{n-1} H_D)$, where
\[
H_\alpha=-\frac{d^2}{dx^2} + \frac{\alpha}{n}\,\sum_{l\in\ZZ} \delta(x-\pi l)
\]
and $H_D$ is the Laplace operator in $L^2(\RR)$ acting on functions
satisfying the Dirichlet boundary conditions at the points $\pi l$, $l\in \ZZ$.
Therefore, the spectrum of $H$ consists of the spectrum of $H_\alpha$
and of the infinitely degenerate eigenvalues $m^2$, $m\in \NN$.

For $\alpha\ne 0$, the spectrum of $H_\alpha$
consists of values $k^2$ satisfying the Kronig-Penney equation
$ \big|\cos \pi k+\alpha/(2nk)\, \sin \pi k\big|\le 1 $, $\Im k\ge 0$, 
and the band edges are given by the values $k^2$ with
$\cos \pi k+\alpha/(2nk)\, \sin \pi k=\pm1$,
see \cite[Theorem III.2.3.1]{AGHH}. In particular,
the Dirichlet eigenvalues are situated on the band edges.

\end{example}

\begin{example}[Periodic Kirchhoff coupling]
Let us emphasize a particular case of the previous example.
If $\alpha=0$ (Kirchhoff couplings), then $H_\alpha$ is just the free Laplacian.
The spectrum of the initial operator $H$, i.e. of channels
periodically coupled by the Kirchhoff boundary conditions,
consists of the semiaxis $[0,+\infty)$ and embedded Dirichlet eigenvalues $m^2$,
$m=1,2,\dots$.
\end{example}

The existence of eigenvalues in the spectrum of a periodic problem on the graph
in our toy situation is connected closely with the reducibility of the boundary conditions.
Nevertheless, such effects appear in much more general structures~\cite{Ca}.
It is known that a periodic graph can have eigenvalues only in the case
of compactly supported solutions~\cite{Ku2}. The existence
of such solutions is possible only in the case of the so-called analytically
disjoint couplings, which can produce even stronger spectral effects~\cite{OP}.

\section*{Acknowledgments}
The author thanks Vladimir Geyler, Volodymyr Mikhailets, Olaf Post, and Nader Yeganefar for stimulating discussions and valuable remarks. The paper was considerably improved following the comments of one of the anonymous referees. The author is indebted him very much. The work was partially supported by the Sonderforschungsbereich \mbox{``Raum~$\bullet$~Zeit~$\bullet$~Materie''}(SFB 647, Berlin), INTAS, and the program of cooperation between the Deutsche Forschungsgemeinschaft and the Russian Academy of Sciences.


\end{document}